\documentclass{article}

\usepackage{arxiv}
\usepackage[numbers,sort&compress]{natbib}
\usepackage{chemformula} 
\usepackage[T1]{fontenc} 
\usepackage{graphicx}
\usepackage{subcaption}
\usepackage{cleveref}
\usepackage{color}
\usepackage{adjustbox}
\usepackage{makecell}
\usepackage{soul}
\usepackage{booktabs}
\usepackage{import}

\title{A plate-type condenser platform with engineered wettability for space applications}

\author{Tibin M. Thomas\\
	Department of Mechanical Engineering\\
	Indian Institute of Technology Madras\\
	Chennai 600036, India \\
	\texttt{me17d042@smail.iitm.ac.in} \\
	\And
	Pallab Sinha Mahapatra \\
	Department of Mechanical Engineering\\
	Indian Institute of Technology Madras\\
	Chennai 600036, India \\
	\texttt{pallab@iitm.ac.in} 
}
\renewcommand{\shorttitle}{A plate-type condenser platform}

\begin{document}
\maketitle

\begin{abstract}
Vapor condensation is extensively used in applications that demand the exchange of a substantial amount of heat energy or the vapor-liquid phase conversion. In conventional condensers, the condensate removal from a subcooled surface is caused by gravity force. This restricts the use of such condensers in space applications or in horizontal orientations. The current study demonstrates proof-of-concept of a novel plate-type condenser platform for passively removing condensate from a horizontally oriented surface to the surrounded wicking reservoir without gravity. The condensing surface is engineered with patterned wettabilities, which enables the continuous migration of condensate from the inner region of the condenser surface to the side edges via surface energy gradient. The surrounding wicking reservoir facilitates the continuous absorption of condensate from the side edges. The condensation dynamics on different substrates with patterned wettabilities are investigated, and their condensation heat transfer performance is compared. The continuous migration of condensate drops from a superhydrophobic to a superhydrophilic area can rejuvenate the nucleation sites in the superhydrophobic area, resulting in increased heat transport. We can use the condenser design with engineered wettability mentioned above for temperature and humidity management applications in space. 
\end{abstract}


\section{Introduction}
Condensation is an inevitable process in energy conversion \cite{li2019organic}, environmental control \cite{perez2011review}, water harvesting \cite{nioras2021different}, desalination \cite{nassrullah2020energy}, and thermal management systems \cite{song2021multifunctional}. The overall efficiency of these systems has a greater dependence on the condensation process, and the performance of the condensation process is reliant on the mode of condensation, nucleation rate, and condensate removal rate \cite{schmidt1930condensation, tanaka1979further}. These parameters are influenced by the surface wettability, which can be engineered by both physical and chemical modifications \cite{attinger2014surface,cho2016nanoengineered,tran2022ultrafast,wilke2020polymer}. The gravitational force plays a vital role in facilitating the removal of condensate from a surface, which is necessary for the continuous functioning of conventional condensers at optimum performance. The transport of condensate over the surface and the removal of condensate become challenging when a condenser surface is placed in a microgravity environment (space) or horizontal orientation. In space applications, devices such as heat pipes \cite{mangini2017hybrid, chatterjee2011constrained}, porous media based condensers \cite{hasan2006conceptual}, and other complex condensing heat exchanger systems consisting of slurper bars and a rotary separator \cite{hauser2002condensing, thomas2014condensing} are used as a replacement for conventional condenser devices. Despite the fact that the wicking process ensures condensate transport via capillary forces, the fundamental disadvantage of these devices is their poor heat transfer performance due to film formation on the condensing area. Because the heat transfer performance of the filmwise mode of condensation during pure vapour conditions is one order of magnitude lower than the dropwise mode \cite{schmidt1930condensation}. It is challenging to design a plate type condenser with superior performance that can be employed in space or in a horizontal orientation with either a dropwise mode or a mix of both filmwise and dropwise modes. The various techniques used for the self-transport of liquid on an open surface without gravity aid are important for the development of condensers that can efficiently remove the condensate from the surface for space application.    

Because of the exciting opportunities in micro-fluidics, lab-on-a-chip, condensation, and spray cooling applications, self-transport of liquid on an open surface has attracted a lot of attention in the last decade \cite{lowrey2021survey,dai2020directional,thomas2021droplet,sinha2022patterning}. Passive liquid transport on an open surface without gravity aid was accomplished through engineered surface modification techniques such as wettability gradient \cite{chowdhury2019self}, superhydrophilic wedge surrounded by a hydrophobic boundary \cite{sen2018scaling}, topological liquid diode \cite{li2017topological}, charge gradient \cite{sun2019surface}, and so on. On a wettability gradient surface, the difference in the Laplace pressure between two opposite ends drives the droplet from the high contact angle region to the low contact angle region \cite{chowdhury2019self}. Likewise, liquid transport occurs from the narrower end to the wider end of a superhydrophilic wedge track due to the Laplace pressure difference between the front and back end \cite{sen2018scaling,stamatopoulos2020droplet}. The topological structure on the surface of a topological liquid diode breaks the contact line pinning of the drop at the advancing edge while simultaneously imparting a strong pinning force at the receding edge. This converts excess surface energy to kinetic energy, propelling the droplets from receding to advancing end \cite{li2017topological}. On a charge gradient surface, droplets transport from a higher charge density region to a lower charge density region \cite{sun2019surface,zhang2021charge}.

In the presence of gravity, the condensate is removed from the vertically oriented surface when the size of the condensate reaches the capillary length scale, such that the gravity force of the condensate exceeds the surface tension forces \cite{schmidt1930condensation}. Condensate droplets can also be removed at a size smaller than the capillary length scale via coalescence-induced droplet jumping from non-wettable surfaces \cite{boreyko2009self, aili2016unidirectional}, periodic wicking structures from a hydrophilic surface \cite{preston2018gravitationally,cheng2021rapid}, and the application of an intermittent electrostatic potential to a hydrophilic surface \cite{bahadur2007electrowetting}. In addition, various techniques of self-transport of liquid drops have been used in the condensation process to achieve directional transport of condensates or to improve overall condensation performance. On a surface with radially graded wettability, condensate was observed migrating from the central region of a substrate to the outer region \cite{daniel2001fast,macner2014condensation}. Many researchers have also proposed bio-inspired structural graded structures that mimic spider silk, desert beetles, cactus spines, pitcher plants, and other natural materials for water harvesting from humid air via condensation and directional transport of condensate liquid \cite{sharma2018gladiolus,zhang2017bioinspired,park2016condensation, hou2015recurrent}. When a superhydrophobic surface is infused with high viscous oil, condensate droplets can be seen levitating. Capillary force causes a meniscus to form over a condensate drop on an oil-infused surface. The stretching of the oil layer caused by droplet growth results in the droplets levitating \cite{sun2019microdroplet}. Different types of surfaces with patterned wettability have also been proposed in the literature for directional condensate transport and to utilise the advantages of both hydrophilicity and hydrophobicity during condensation \cite{ghosh2014enhancing,mahapatra2016key, nioras2022atmospheric, tang2021design, koukoravas2020experimental}. 

Previous research on condensation on a horizontally oriented substrate focused on either condensation mechanisms like nucleation, drop growth, and drop-coalescence, or passive droplet movement on the substrate with a radially smooth wettability gradient \cite{daniel2001fast,macner2014condensation} or a charge density gradient \cite{zhang2021charge}. Nonetheless, it is yet unclear how to employ these substrates as condenser systems and how to continuously remove the migrating droplets from the substrate in a microgravity environment. Additionally, previous studies on the condensation process on a substrate with patterned wettability and the effectiveness of their condensation heat transfer primarily looked at vertically oriented or inclined substrates \cite{ghosh2014enhancing,mahapatra2016key,derby2014flow}. In such cases, the gravity force aids the removal of condensate from the substrate irrespective of the design of the various patterned wettabilities. On the other hand, when condensation occurs on a horizontal surface or a condenser surface kept in a microgravity environment, the condensate spreads and accumulates regardless of the substrate wettability. In such cases, the removal of condensate from the surface is challenging and condensate accumulates on the substrate with time. To avoid liquid accumulation, condensate must be continuously removed from the surface using either passive or active techniques.

The current study proposes proof-of-concept for a plate-type condenser platform with patterned wettability as well as a novel method for the continuous removal of liquid formed by condensation on a horizontally oriented surface or the condenser surface in a microgravity environment via a wicking reservoir. A substrate with patterned wettability was designed to allow the passive transport of condensate droplets without external force or gravity. The gradient in surface energies propels the condensate formed on the substrate to the surface's outer edges. The propelled condensate from the outer edges is absorbed by the wicking reservoir, which is surrounded by the substrate. Wherein, the wicking reservoir is not attached to the substrate, and a tiny gap is maintained between them to prevent heat conduction. The mechanism by which condensate is transported from the substrate to the wicking reservoir is explored. The key features of this study are,

\begin{itemize}

 \item Different patterned wettable surfaces were used and a design is proposed (pattern P4) as a best condensing surface for the condensate transport. The proposed pattern P4 and the passive droplet transport process is novel. This design showed better condensate collection during experiments and the transport mechanism is explained through simulations.
 
\item A new approach to the experimental characterization of the condensation rate from a horizontally oriented substrate by a wicking reservoir was established in the present study.  
  
\item Although, a lot of data on condensation heat transfer rate were available on literature, however to the best of our knowledge, no data are available with condensation on horizontal surface in presence of NCG.     
\end{itemize}

\section{Material and Methods}
\subsection{Fabrication of the substrate with patterned wettability and characterization}
The substrate material was an aluminium alloy of grade 6061 with dimensions of 40 $mm$$\times$ 40 $mm$$\times$ 3 $mm$. The substrate was first ultrasonically cleaned for 10 minutes in a solution of deionized (DI) water, ethanol, and acetone, and then dried with $N_2$ gas. Further, the substrate was microtextured by dipping in 3$M$ $HCl$ (Merck-EMPLURA grade) for 5 \textit{minutes} and washed thoroughly with DI water. The microtextured substrate was immersed in DI water at a temperature of 100 $^\circ C$ for 30 \textit{minutes} to grow the boehmite nanostructures above the microstructures \cite{vedder1969aluminum}. These procedures modified the polished aluminum substrate to a hierarchical rough superhydrophilic surface with a contact angle of below 5$^\circ$. Further, the superhydrophilic substrate was functionalized with 0.5 \% vol/vol 1H,1H,2H,2H-Perfluoro-octyl-triethoxysilane (PFOTS) in ethanol solution by dip-coating for 3 \textit{hours}, followed by drying in a laboratory environment for 12 \textit{hours} \cite{thomas2021condensation}. This procedure transformed the wettability of the substrate from superhydrophilic (SHL) to superhydrophobic (SHB). To fabricate the substrate with patterned wettability, the uniformly coated hydrophobic layer has been selectively removed by laser ablation. A 60 $W$ $CO_2$ laser (Universal Laser VLS3.60) having a spot size of 30 $\mu m$ was used for the laser ablation. The power setting of the laser was 95 \% of the total power and the speed setting was 5 \% of the total speed. After the laser ablation, the areas that were exposed to the laser changed into SHL, while the areas that weren't exposed to the laser remained as SHB \cite{thomas2021droplet}.

The morphology of the modified hierarchical surface textures was analyzed using a high-resolution scanning electron microscope (Apreo- Thermofisher) after the gold sputtering operation for 30 \textit{sec}. The surface roughness was measured using a surface profilometer (Nanomap 1000WLI). The dynamic contact angles (advancing and receding) were measured by injecting/drawing out the DI water at a flow rate of 0.2 $\mu L/sec$ to/from a sessile water drop of volume 5 $\mu L$ using a syringe pump \cite{gao2009wetting}. 

\subsection{Fabrication of the wicking reservoir}

\begin{figure}[t]
    \centering
    \includegraphics[width=0.8\textwidth]{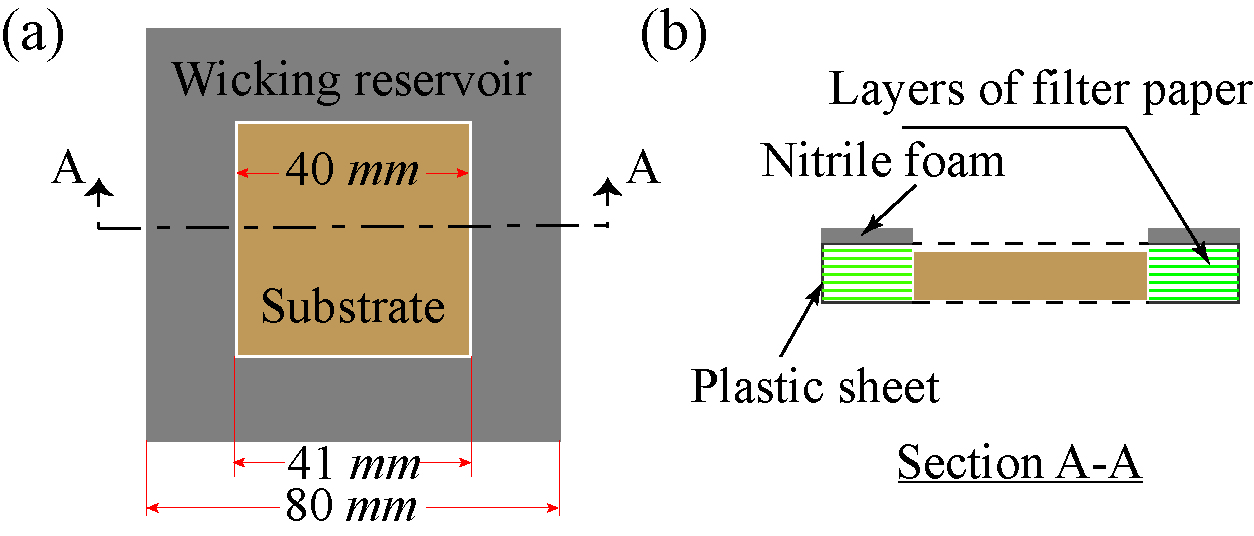}
    \caption{Design of the wicking reservoir (a) Illustration of wicking reservoir (gray colour) placed surrounded to the condensing substrate. To prevent heat transfer losses due to conduction between the wicking reservoir and the substrate, a gap of 0.5 $mm$ was maintained. (b) Sectional view of wicking reservoir, comprising of layers of filter papers, plastic layer covering and nitrile foam insulation.}
    \label{fig:w_res}
\end{figure}
A wicking reservoir was used to passively remove the condensate from a horizontal substrate. The wicking reservoir is made up of multiple layers of Whatman filter papers of Grade 1 as interior layers and a thin plastic sheet as top and bottom layers, as shown in Fig. \ref{fig:w_res}. Each layer of filter paper was cut according to the dimensions mentioned in Fig. \ref{fig:w_res}a using a $CO_2$ laser at a power setting of 5 \% of total power and 3 \% of total speed. The laser-cut papers were then stacked, and the top and bottom sides of the stack were covered with a laser-cut plastic sheet of the same dimension using double-sided adhesive tape. To prevent mass loss of condensates collected from the substrate, the lateral sides of the wicking reservoir were covered with plastic tape. The top surface of the wicking reservoir was insulated with 3 $mm$ of thick nitrile foam to prevent water vapour condensation on the wicking reservoir (see Fig. \ref{fig:w_res}b).

\subsection{Experimental procedures}

\begin{figure}[t]
    \centering
    \includegraphics[width=\textwidth]{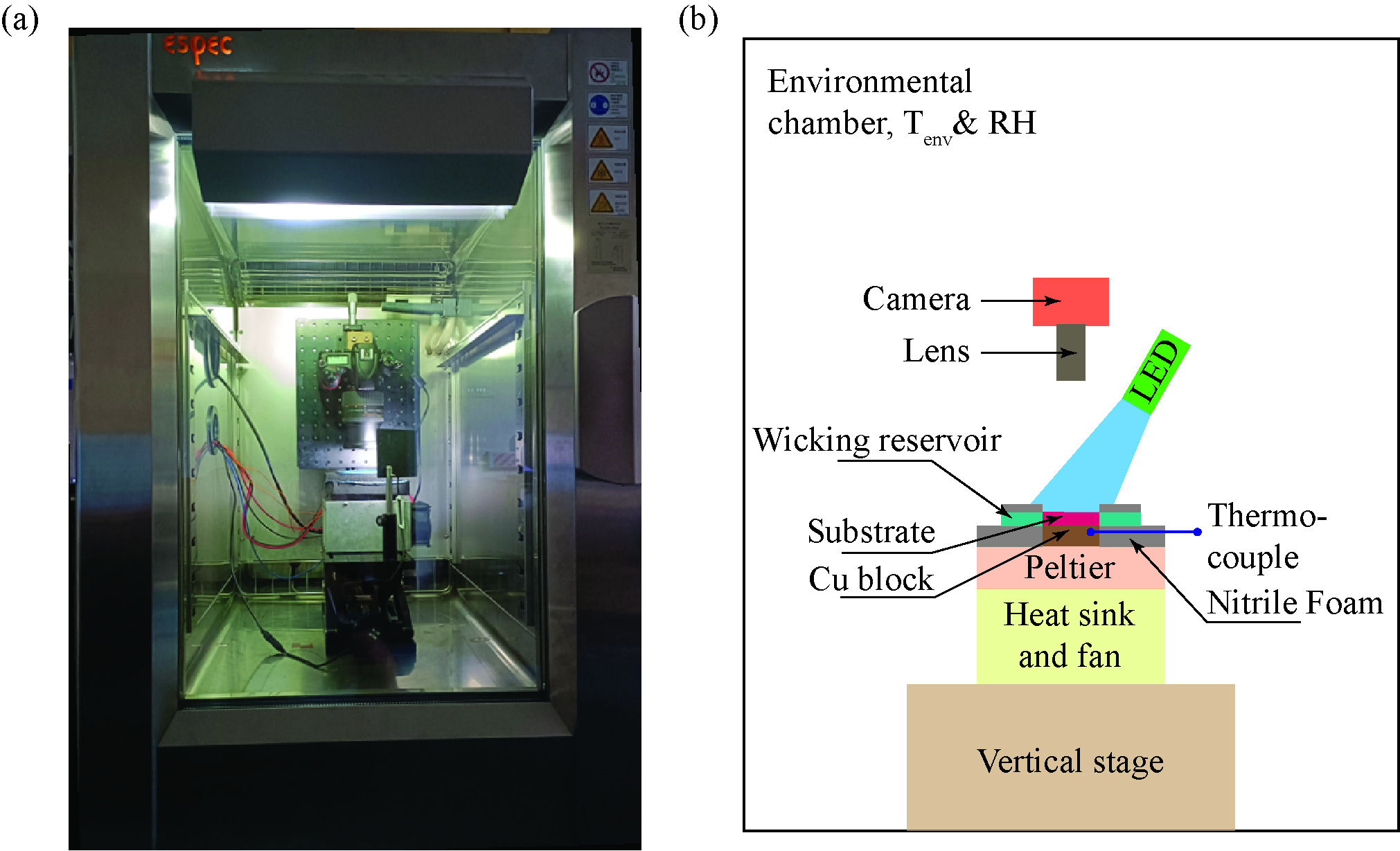}
    \caption{Experimental setup for the condensation experiments on the horizontally oriented substrate with patterned wettability. (a) Snapshot and, (b) schematic representation. }
    \label{fig:setup}
\end{figure}
The condensation experiments on horizontally oriented surfaces were performed in an environmental chamber (Espec-PR2J), as shown in Fig. \ref{fig:setup}. A peltier (TeTech-CP061HT) was used to cool the test substrate, which was placed on a vertical stage in a horizontal orientation. A copper block of dimension 40 $mm \times$ 40 $mm \times$ 10 $mm$ was attached to the middle of the peltier cold plate using double-sided thermal conductive tape (3M Tape) of a thickness of 0.25 $mm$. The remaining area of the peltier top plate was covered with a 10 $mm$ thick nitrile foam as thermal insulation. Then, the substrate was attached above the copper block using thermal conductive double-sided tape (3M Tape of 0.25 $mm$). A wicking reservoir was placed around the substrate to wick the condensate from the substrate. To prevent heat transfer loss by conduction, direct contact between the wicking reservoir and the substrate was avoided, and a gap of 0.5 $mm$ was kept between the substrate and the wicking reservoir (see Fig. \ref{fig:w_res}). Thus, in order to initiate wicking, the liquid-vapor interface of the condensate must crosses the gap and reach up to the side edges of the wicking reservoir. The overall condensation performance of the different substrates was measured from the weight difference of the wicking reservoir before and after the condensation experiments in a controlled environment for a duration of 1-2 \textit{hours}. The mass of the wicking reservoir was measured with a micro-balance (Ohaus-SPX622) of 10 $mg$ accuracy. A sheathed T-type thermocouple (Tempsens) of diameter 0.6 $mm$ was inserted through the lateral side of the substrate by making a drilled hole of diameter 0.8 $mm$. The temperature of the substrate was measured using a data acquisition system (Keysight-970A) at a recording rate of 1 \textit{data/minute}. The time lapse images of the condensation dynamics on the substrate were recorded at every \textit{minutes} using a DSLR camera (Nikon-D750) with a zoom lens (24-85 $mm$ Nikon lens), at a focal length of 85 $mm$.   

\section{Results and Discussions}
\subsection{Surface morphology and engineered wettability}

\begin{figure}[t]
    \centering
    \includegraphics[width=0.9\linewidth]{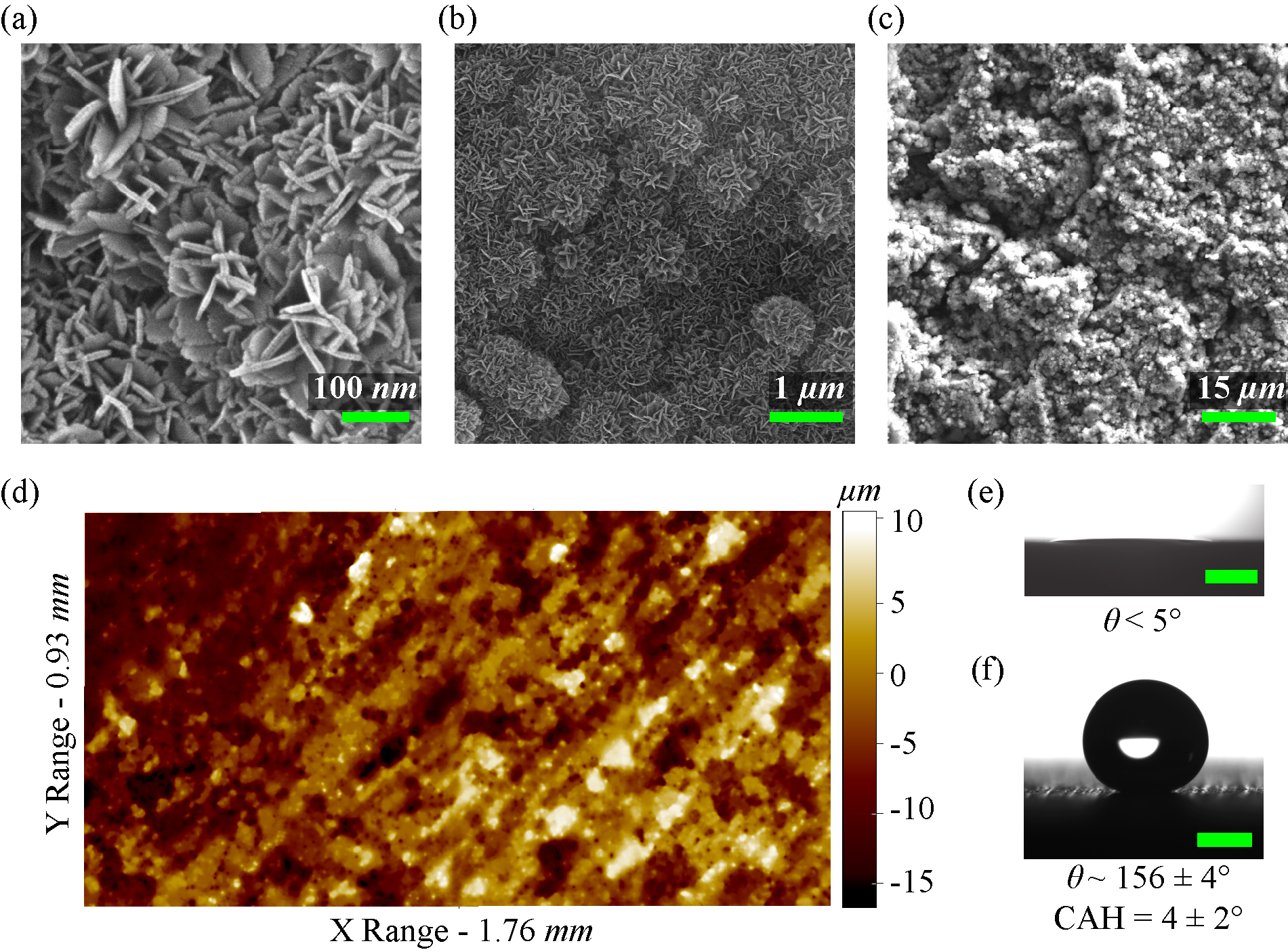}
    \caption{Surface morphology and wettability. (a-c) SEM, and (d) profilometer micrographs of the SHB surface. Contact angle images of the water drop (e) before and, (f) after hydrophobic functionalization on textured aluminum substrate. The scale bars in the contact angle images are 1 $mm$.}
    \label{fig:sem}
\end{figure}

Figure \ref{fig:sem}(a-c) shows the SEM images of the fabricated SHB surface at different resolutions. It is evident from the SEM micrographs that the rationally modified surface consists of $Al_2O_3$ blade-shaped nanostructures and flower-shaped microstructures \cite{thomas2021condensation}. The length and width of the blade shaped nanostructures were in the range of 100-300 $nm$ and 15-30 $nm$ respectively. The size of the flower-shaped microstructures was in the range of 0.5-1 $\mu m$. The physical structure of the SHB surface is identical to the surface morphology of the textured surface before the silane functionalization, since the self-assembled monolayer could only make a layer of negligible thickness of 30 $\AA$ \cite{yang2006dropwise}. The average macroscopic roughness of the textured surface was $\sim$4.2 $\mu m$. The textures on the modified rough surface were random, and the peak height and valley depth were in the order of $\sim$10 $\mu m$ and $\sim$15 $\mu m$ respectively (see Fig. \ref{fig:sem}d). The contact angle of the sessile water droplet before the hydrophobic coating was nearly zero. The hydrophobic coating changed the surface wettability of the textured surface to superhydrophobicity, as shown in Fig. \ref{fig:sem}(e-f). The magnitudes of the advancing ($\theta_A$) and receding ($\theta_R$) contact angles on the SHB surface were $159\pm2^\circ$ and $155\pm2^\circ$ respectively. The contact angle hysteresis (CAH) is defined as the difference between advancing and receding contact angles, and the magnitude of the CAH for the SHB surface was $4\pm2^\circ$. The wettability patterns on the substrate were fabricated by selective etching of the hydrophobic coating using a $CO_2$ laser. As a result, the laser exposed area on the substrate transitioned to superhydrophilic with a contact angle of below 5$^\circ$.   

\begin{figure}
    \centering
    \includegraphics[width=0.9\linewidth]{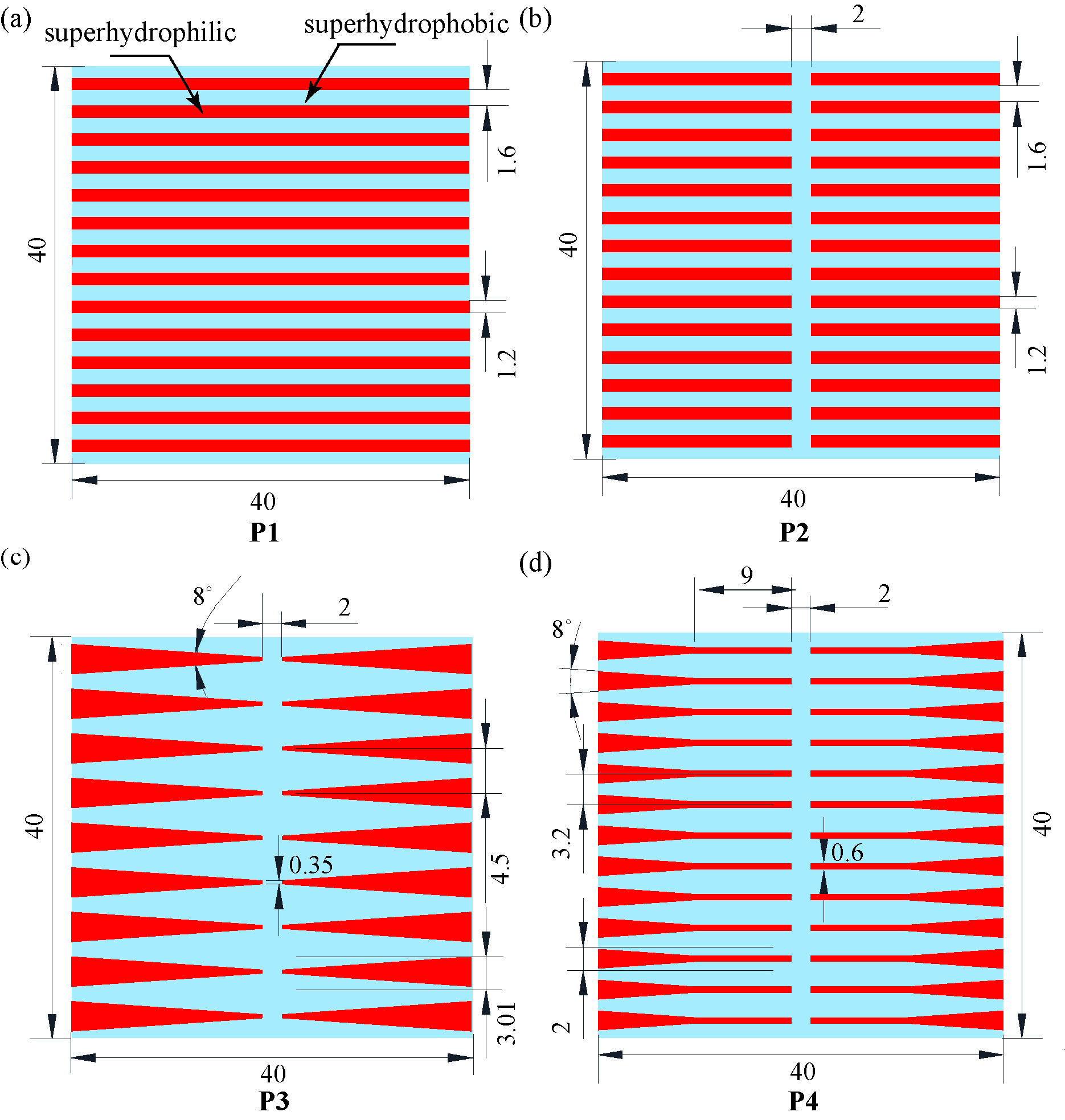}
    \caption{A plate type condenser substrate with patterned wettability comprising of SHL and SHB wettability regions (a) Pattern P1 - rectangular strip: full, (b) Pattern P2 - rectangular strip: half, (c) Pattern P3 - wedge, (d) Pattern P4 - hybrid (wedge + rectangular strip). The red shaded region represents the SHL area and the blue region represents the SHB area. Here, the SHB region promotes the dropwise mode of condensation and the SHL region promotes the filmwise mode of condensation. The SHL tracks facilitate the transport of condensate from the inner region of the substrate to the side edge region. The units of all dimensions mentioned in this figure are in $mm$.}
    \label{fig:pattern}
\end{figure} 

Figure \ref{fig:pattern} illustrates the design of the different substrates with patterned wettabilities used in the present study. The pattern P1 consists of fourteen parallel SHL rectangular strips with a width of 1.2 $mm$ and a length of 40 $mm$ laid on a SHB background (see Fig. \ref{fig:pattern}a). Wherein, the spacing between the wettability transition lines of two consecutive SHL strips is set at 1.6 $mm$. For pattern P2, every SHL track in the pattern P1 is divided from the centre such that each of these tracks is converted to two rectangular tracks, as shown in Fig. \ref{fig:pattern}b. Hence, the total number of SHL tracks became twenty-eight in the pattern P2 with a strip length of 19 $mm$. The central spacing between two in-line SHL strips is set at 2 $mm$. In pattern P3, rectangular strips of pattern P2 were replaced with wedge-shaped tracks having a wedge angle of $8^\circ$ (see Fig. \ref{fig:pattern}c). Wherein, the width of the wedge at its widest end was 3 $mm$, the narrower end was 0.35 $mm$ and the pitch between two consecutive parallel wedges was 4.5 $mm$. On pattern P3, the distance between the wettability transition lines of two consecutive SHL wedges ranges from 4.15 $mm$ at the narrower end to 1.5 $mm$ at the wider end. The wedge angle of 8$^\circ$ is chosen in the design such that successful transport of a drop with a diameter greater than 1.8 $mm$ (equal to the departure diameter for a hydrophobic surface in vertical orientation \cite{wen2017hierarchical}) from the superhydrophobic area to the surrounding wicking reservoir is achievable once the drop has merged with the liquid at the narrower end of the wedge. In our earlier study, we described the rationale behind selecting the various parameters in the proposed pattern of wettability. \cite{thomas2021droplet} The total number of SHL tracks in the pattern P3 was reduced to eighteen compared to pattern P2 since the width of the tracks in the pattern P3 was increased to 3.01 $mm$ compared to 1.2 $mm$. To increase the total number of SHL tracks, pattern P4 was designed by combining patterns P2 and P3. The various design parameters for pattern P4 were selected so that the total fractional area of the superhydrophilic track of the surface was kept in the range between 30--35\%, like in pattern P3. Each SHL track in pattern P4 consists of a shorter wedge compared to the wedge in pattern P3, and the narrower end of the wedge was connected with a rectangular strip of width 0.6 $mm$ and a length of 9 $mm$, as shown in Fig. \ref{fig:pattern}d. The total number of SHL tracks on Pattern P3 increased to twenty-six. The width of the wider end of the wedge was 2 $mm$, the narrower end of the wedge was 0.6 $mm$, and the pitch between the two consecutive tracks was 3.2 $mm$. As a result, the distance between the wettability transition lines of two consecutive SHL tracks on pattern P4 was shortened in comparison to pattern P3, ranging from 2.6 $mm$ at the narrower end to 1.2 $mm$ at the wider end.     

\subsection{Condensation observation}
\begin{figure}[h]
    \centering
    \includegraphics[width=\linewidth]{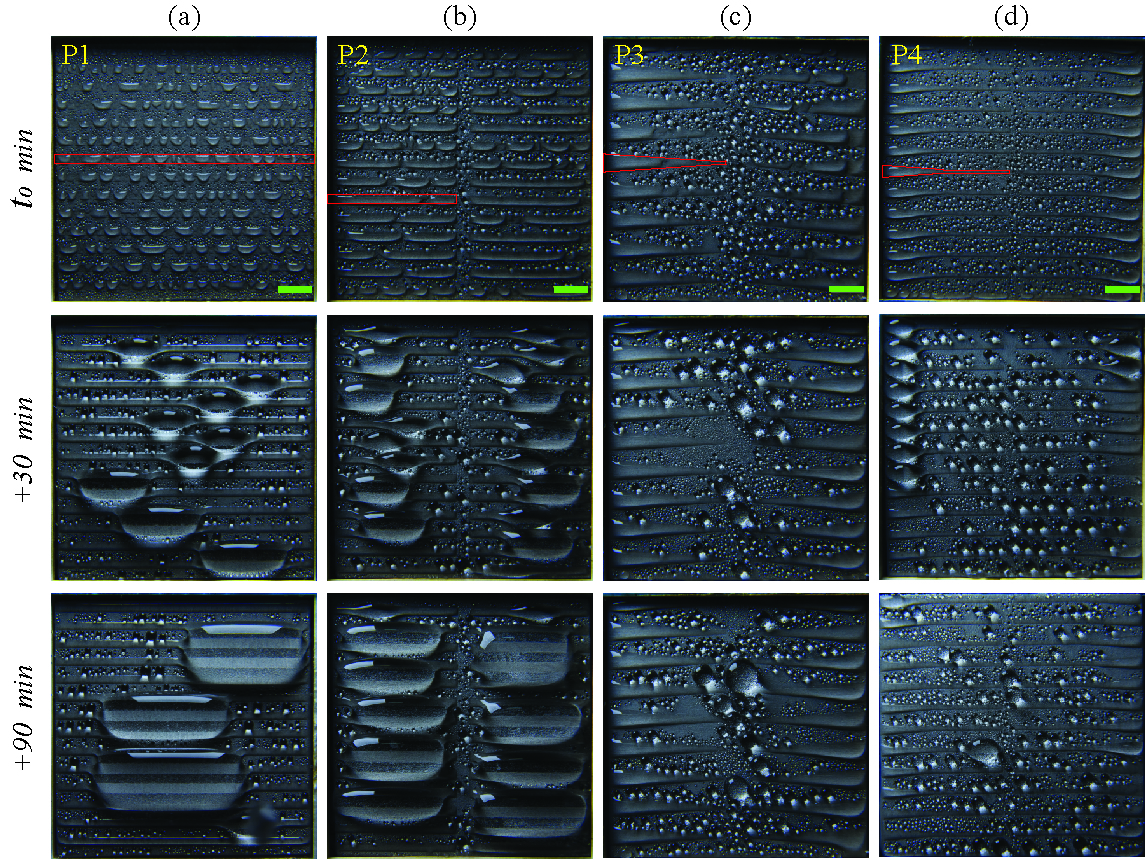}
    \caption{Macroscale condensation snapshots of different substrates with patterned wettability at different times. The red-coloured contour represents the shape of the SHL track in each patterned substrate. $t_0$ denotes the time when the temperature of the substrate reaches $T_s$. The scale bar is 5 $mm$.}
    \label{fig:snaps}
\end{figure}
Condensation is a complex phenomenon that consists of various processes such as drop nucleation, droplet growth, and coalescence \cite{tanaka1979further}. Typically, condensate that forms on a surface drains away when the gravity force of the condensate becomes higher than the force of surface adhesion. However, in a microgravity environment, the generated condensate on a homogeneous surface coalesces and accumulates over the surface as multiple bulges, further transitioning to a thick film. The removal of condensate from a surface before the formation of a larger sized liquid bulge is necessary for the continuous operation of a condenser at a desired heat transfer rate. The wicking reservoir of the proposed platform helps to remove the condensate from the substrate. The liquid-vapor interface of the condensate must overcome the gap between the substrate side edges and the wicking reservoir in order to start wicking. The condensate transport mechanism and its performance on four different substrates with patterned wettability in a controlled humid air environment are compared in this study. Figure \ref{fig:snaps} depicts the snap-shots of the condensation phenomena on the horizontally oriented substrates having different patterned wettabilities at an interval of 30 \textit{minutes.} The experiments were carried out in an environmental chamber with a temperature ($T_{env}$) of 40 $^\circ C$, a relative humidity (RH) of 80 \% and a substrate temperature ($T_s$) of 4$\pm$1 $^\circ C$. At the initial stage of condensation, the substrates showed a dropwise mode of condensation on the SHB area and a filmwise mode of condensation on the SHL area. On pattern P1, the amount of condensate liquid increased in the rectangular SHL tracks and formed a liquid bulge with time since the condensate was not removed from the substrate. This liquid bulge grew with time, and the liquid interface of the bulge coalesced with either the neighbouring bulge or the neighbouring SHL rectangular track. After a few more coalescences, the liquid bulges over the centre region of the substrate and floods (see Fig. \ref{fig:snaps}a). These liquid bulges prevent the effective spread of condensate through the SHL tracks and resist the continuous removal of condensate through the substrate side edges by wicking. The formation of larger sized liquid bulges was also observed on the substrate with pattern P2. The total volume and the total number of bulges were found to be higher on pattern P2 than on pattern P1. The central SHB area on pattern P2 avoids the merging of liquid bulges formed on the SHL tracks found on the opposite sides, as shown in Fig. \ref{fig:snaps}b. Interestingly, flooding of the condensate was not observed on patterns P3 and P4 (see Fig. \ref{fig:snaps}c and d). This indicates that the nucleated condensate was being removed continuously from these substrates with patterned wettability P3 and P4. 

\begin{figure}
    \centering
    \includegraphics[width=\linewidth]{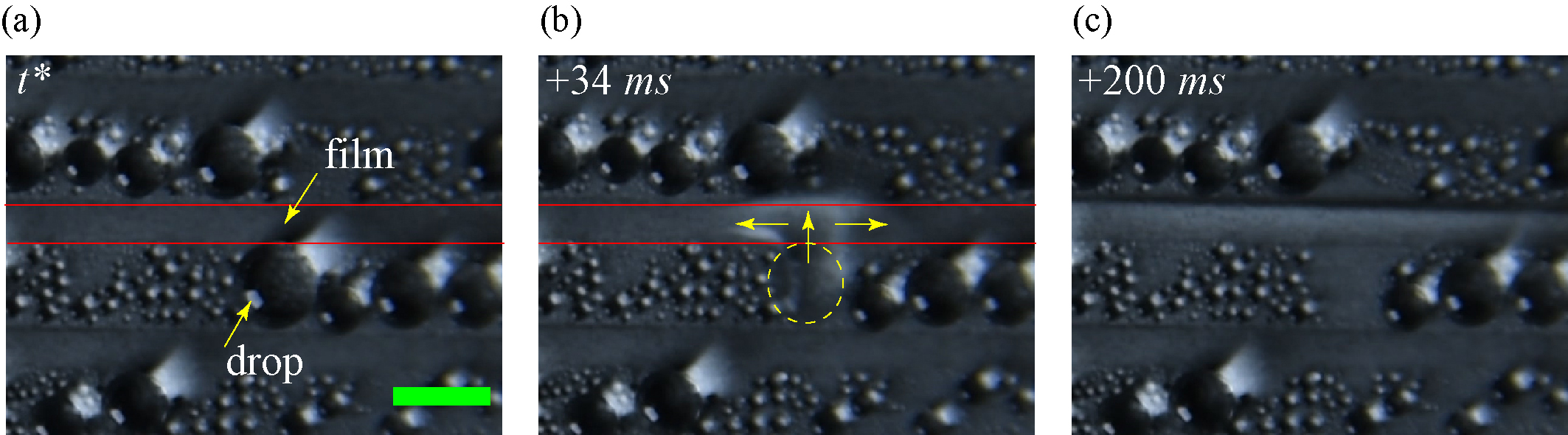}
    \caption{The coalescence between drop and film causes the capillary-driven pumping of condensate from the SHB region to the SHL region on pattern P1 and P2. The red solid line represents the border line between the SHB and SHL areas. The yellow arrow line represents the liquid transport from drop to film. The scale bar is 3 $mm$.}
    \label{fig:pump}
\end{figure}
Condensate appeared as distinct drops with a size ranging from several nanometers to millimetres in the SHB area. Whereas, condensate was observed in the form of a liquid film on the SHL tracks. The condensate drops from the SHB region were passively transported to the SHL tracks by coalescence driven capillary pumping \cite{derby2014flow, mahapatra2016key}. The nucleated drop in the SHB region grew with time, and it coalesced with the nearest condensate film in the SHL region once the liquid interface of the drop was in contact with the wettability transition line, as shown in Fig. \ref{fig:pump}. This liquid transport occurs due to the difference in the Laplace pressure acting at the liquid-vapor interface between the drop and the film. The higher interfacial Laplace pressure of the drop compared to the liquid film having an infinite radius of curvature causes the transport of liquid from the drop to the film in the SHL track. After the coalescence, the liquid-vapor interface of the liquid film in the SHL region crosses the wettability transition line (see Fig. \ref{fig:pump}c) due to inertia, and finally the liquid volume from the drop is spread out over the SHL track. 

\begin{figure}
    \centering
    \includegraphics[width=\linewidth]{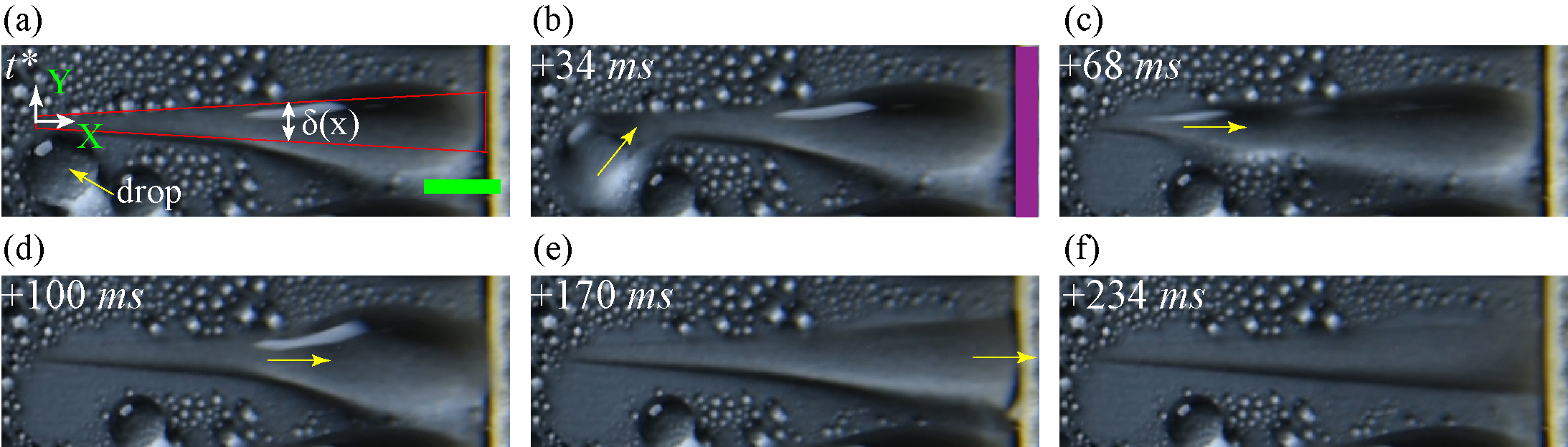}
    \caption{An illustration of the passive transport of condensate through an SHL wedge and the transport of condensate from substrate to wicking reservoir in the pattern P3. The red solid line represents the border line between the SHB and SHL areas. The yellow arrow line represents the direction of liquid transport. The violet-coloured shaded region represents the side end of the wicking reservoir. The scale bar is 3 $mm$.}
    \label{fig:wedge}
\end{figure}

Figure \ref{fig:wedge} demonstrates the transport of a liquid drop from the SHB region to the wicking reservoir via the SHL wedge on substrate P3 during condensation. Wherein, the SHL track is completely filled with the condensate before the drop-film coalescence event. The drop from the SHB region is transported to the SHL region once the liquid interface of the drop is in contact with the liquid film present in the SHL track due to capillary pumping \cite{derby2014flow, mahapatra2016key}. Thereafter, the liquid is transported inside the wedge track from the narrower end to the wider end by a Laplace pressure gradient arising due to the curvature difference between the front and back end \cite{chowdhury2019self,sen2018scaling}. The magnitude of the pressure gradient of the liquid bulge in the SHL track is calculated based on scaling arguments \cite{sen2018scaling},
\begin{equation}
    \frac{dP}{dx}\approx 2\sigma_{lv}sin\theta_{avg}\frac{\alpha}{2\delta(x)^2}
\end{equation}
where $\sigma_{lv}$ is the interfacial surface tension, $\theta_{avg}$ is the average of the front and back end apparent contact angles of the liquid bulge, $\alpha$ is the included angle of the wedge track, and $\delta(x)$ is the track width of the wedge at a local position of $x$ from the narrower end in the direction of liquid transport (see Fig. \ref{fig:wedge}a). Wherein, the SHL track is not completely filled with the condensate prior to drop-film coalescence. The transported liquid accumulated in the wider region of the wedge due to the substrate edge resistance. The size of the accumulated liquid bulge inside the SHL track increases with time, and the liquid interface of the bulge merges with the surrounded wicking reservoir. Thereafter, the wicking reservoir absorbs condensate from the substrate by wicking. 

\begin{figure}
    \centering
    \includegraphics[width=\textwidth]{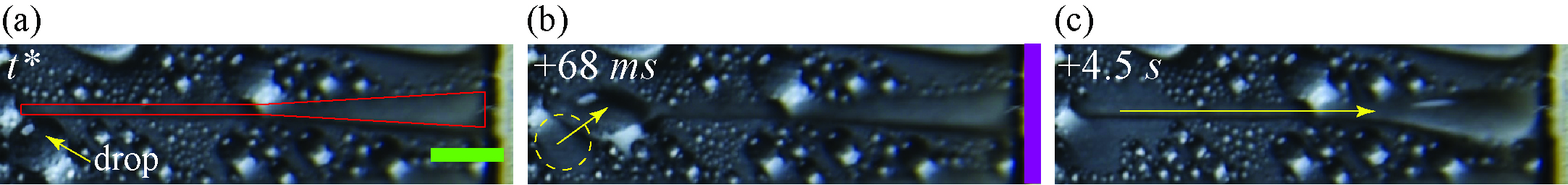}
    \caption{Illustration of liquid transport from the thin rectangular track to the wider end of the wedge in the pattern P4. The red solid line represents the border line between the SHB and SHL areas. The yellow arrow line represents the direction of liquid transport. The violet-coloured shaded region represents the side end of the wicking reservoir. The scale bar is 3 $mm$.}
    \label{fig:hybrid}
\end{figure}

In pattern P4, the hybrid SHL track was designed by superimposing the wedge pattern on a rectangular striped pattern. The condensate transported to the rectangular stripe region from the SHB area spreads into the rectangular shaped region of the hybrid SHL track with time, as shown in Fig. \ref{fig:hybrid}. The transport of condensate is resisted at the merging region between the rectangular and wedge SHL tracks. The maximum accumulation in the rectangular region drives the transport of condensate from the rectangular region to the narrower region of the wedge. Further, the liquid from the narrower end is transported to the wider end by a difference in Laplace pressure, and the accumulated condensate in the wider region of the wedge is absorbed by the surrounded wicking reservoir, similar to pattern P3. The total volume of the liquid that can be accommodated within a single SHL track on pattern P3 is higher as compared to pattern P4. 

The superhydrophilic tracks in the patterned substrate act as an intermediate reservoir before the condensate is removed to the wicking reservoir placed outside of the condenser plate. The drops from the superhydrophobic region can coalesce with the liquid film at different locations on the substrate. After the coalescence, the area occupied by the coalesced drop in the superhydrophobic region will become dry, and further drop nucleation will occur. The time scale of the liquid transport within the superhydrophilic track is depend on the volume of the condensate occupied in the track and the location of the drop when it connected with the wettability transition line. Condensate wicking from the substrate can occur only when the superhydrophilic track is completely filled with condensate and the liquid interface of the bulge bridges the gap between the substrate and wicking reservoir. The time scale for the transport process of the nucleated drops from the superhydrophobic region to the superhydrophilic region is influenced by the pitch between the superhydrophilic track and the total length of the wettability transition line. The total length of the wettability transition line was $\sim$0.7 $m$, and $\sim$1 $m$ for the patterned surfaces P3 and P4, respectively. Hence, the probability for the occurrence of drop to film coalescence process is higher for pattern P4 than pattern P3 due to 42\% increase in the wettability transition line for the pattern P4.
\subsection{Droplet transport on wettability engineered superhydrophilic tracks}
\begin{figure}[h]
    \centering
    \includegraphics[width=0.9\textwidth]{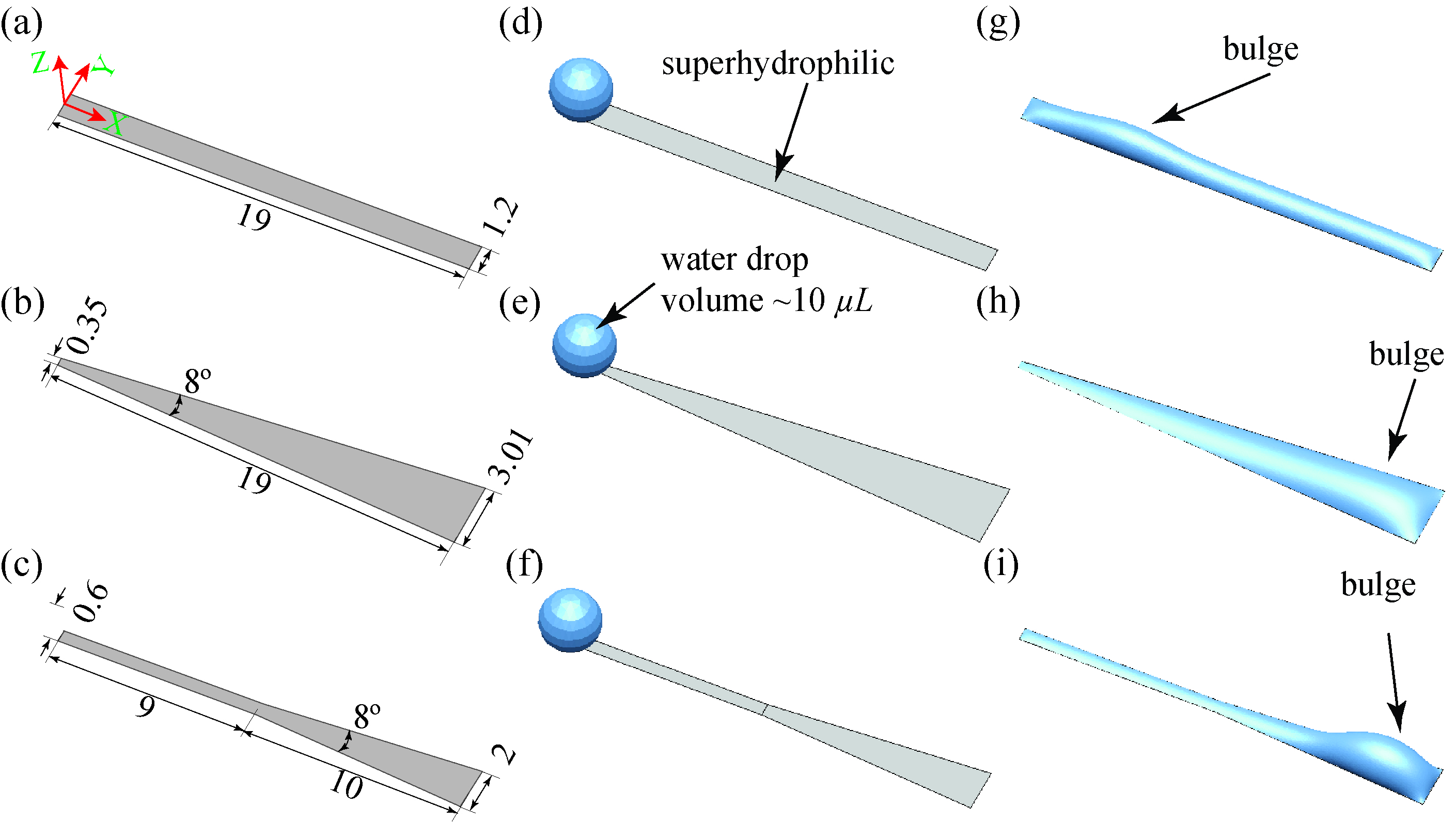}
    \caption{Steady state droplet shape on different patterned superhydrophilic tracks. (a-c) The shape and size of the various superhydrophilic track patterns (Patterns P2, P3, and P4, respectively). The wettability outside of the track is superhydrophobic. All dimensions are in $mm$. (d-f) The droplet's initial condition on a patterned superhydrophilic rectangular track (pattern P2), wedge-shaped track (pattern P3), or combined rectangular and wedge track (pattern P4). (g) The final steady-state shape of the droplet on a rectangular track. The bulge is formed near the initial droplet location. (h) The droplet was transported to the end of the wedge track and the bulge formed near the wider end of the track. (i) The droplet was transported efficiently and a larger-sized bulge formed at the end of the track. The units of all dimensions mentioned in this figure are in $mm$.}
    \label{fig:spread}
\end{figure}
To identify the differences in the droplet transport behaviors on the different patterned wettability surfaces, steady-state simulations are performed using the HyDro droplet simulation solver \cite{matsui2012hybrid}. Matsui \textit{et al.} \cite{matsui2012hybrid} developed HyDro using hybrid energy minimization techniques to identify the equilibrium shape of a droplet placed on a surface. In the simulations, only the transport of a single droplet on smooth patterned wettable surfaces is shown. The actual condensing scenarios are much more complex, with multiple droplet coalescence, droplet-to-film coalescence, and hierarchical surface roughness. To mimic the bulge formation on the superhydrophilic tracks observed during condensation, we have placed a droplet of volume 10 $\mu L$ on different patterned wettable surfaces, as shown in Fig. \ref{fig:spread}. Here the superhydrophilic tracks are laid on a superhydrophobic background. Therefore, when a droplet is placed at the start of the track, the droplet spreads till the end of the track. However, the mechanism of droplet transport in the different tracks or patterns is different. The droplet spreads on the rectangular track (pattern P2) due to the surface's high energy. After the liquid had spread to the end of the track, a bulge formed near the initial droplet location (see Fig. \ref{fig:spread}g). On the wedge track (pattern P3), the liquid spreads due to the net capillary force caused by the difference in curvature between the droplet's front and back \cite{ghosh2014wettability}. As the liquid is pumped due to the capillary force, the bulge formation occurs near the wider end of the track (see Fig. \ref{fig:spread}h). However, the bulge is spread over a longer distance. On the combined rectangular and wedge track (pattern P4), the droplet transport takes place initially on the rectangular track and then reaches the wedge region through the capillary transport on the wedge track. This transport is very efficient, as the liquid can move forward easily through the capillary transport but cannot go back to the rectangular track due to the requirement of extra pressure (working like a natural valve). Hence, the liquid bulge forms near the end of the wedge track (see Fig. \ref{fig:spread}i). The formation of liquid bulges on the various superhydrophilic tracks observed during experiments is identical to the results obtained from steady-state simulations (see Fig. \ref{fig:snaps}, Fig. \ref{fig:wedge}d, and Fig. \ref{fig:hybrid}c).   

\begin{figure}
    \centering
    \includegraphics[width=0.5\textwidth]{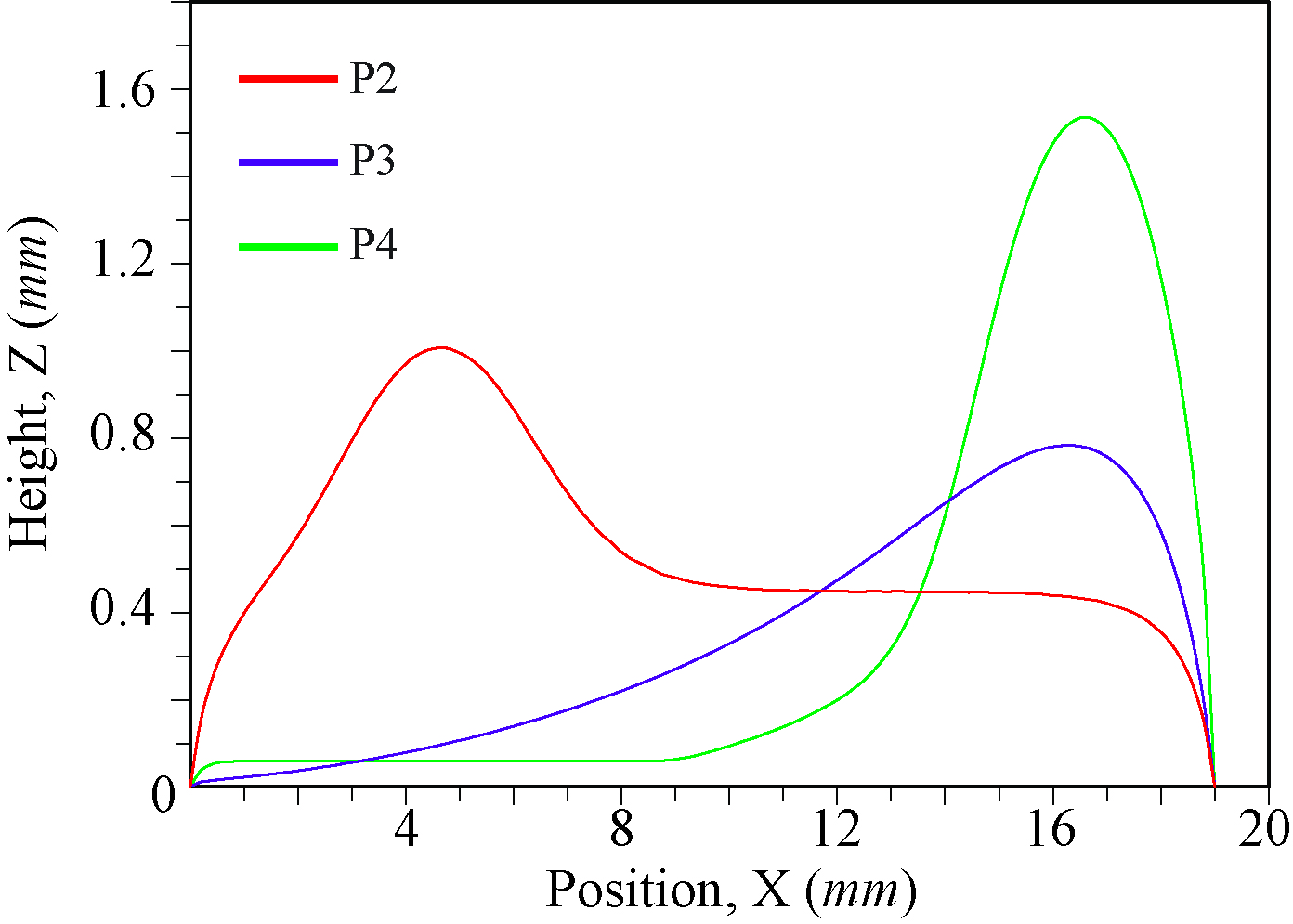}
    \caption{The variation in liquid height at the central plane ($Y=0$) of the different shaped patterns along the track length of the superhydrophilic track. $X=0$ $mm$ represents the initial position of the drop and $X=19$ $mm$ represents the side edge of the substrate.}
    \label{fig:bulge}
\end{figure}

Figure \ref{fig:bulge} depicts the height of the liquid on the central plane ($Y = 0$) of the superhydrophilic track from the start ($X = 0$ $mm$) to the end ($X = 19$ $mm$) position. On pattern P2, the liquid height increases from the initial position and reaches a maximum value at $X=4.75$ $mm$ and further the height reduces till $X=8.5$ $mm$. Thereafter, the liquid height is maintained as constant, and a sharp reduction occurs at the end of the track. For pattern P3, a smooth increase in the liquid height was observed till $X=16.5$ $mm$ and it further reduced at the track end. The liquid height in the wedge region of the pattern P4 first increases upto $X=16.5$ $mm$ and then reduces till the end of the track. Interestingly, in the case of pattern P4, the liquid height in the rectangular region of the track is quite small and the magnitude was $0.06$ $mm$. Thus, the liquid forms a thin film on the rectangular region of the track and a big bulge near the end of the wedge region of the track. Due to the larger bulge at the end of the track, the liquid in the superhydrophilic track of pattern P4 can easily come into contact with the wicking material placed on the sides of the condenser plate.

\subsection{Condensation performance}
\begin{figure}
    \centering
    \includegraphics[width=\linewidth]{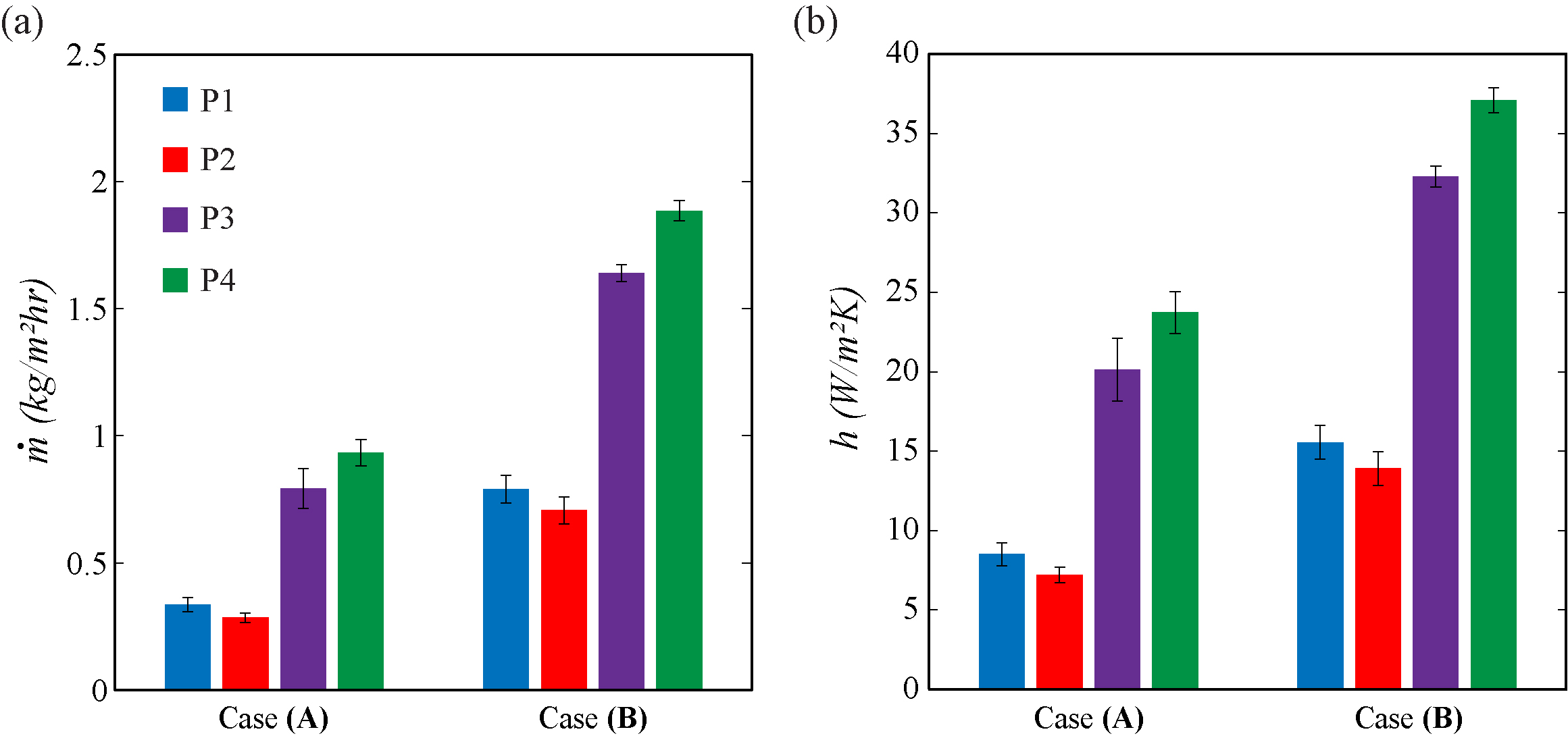}
    \caption{Comparison of condensation performance on different substrates engineered wettability. (a) The average rate of condensate collection by the wicking reservoir on different substrates with patterned wettability. (b) Average heat transfer coefficients of various substrates.}
    \label{fig:mass}
\end{figure}

The condensation experiments on the horizontally oriented substrates were performed under two different conditions: Case (A) $T_{env}=30$ $^\circ C$, RH= 80 \%, and $T_s$= 1.5$\pm$0.5 $^\circ C$, and Case (B) $T_{env}=40$ $^\circ C$, RH= 80 \%, and $T_s$= 4$\pm$1 $^\circ C$. The dew point temperature of Case (A) is $T_{dew}=26.1$ $^\circ C$ and Case (B) is $T_{dew}=35.8$ $^\circ C$. The difference in the mass of the wicking reservoir before and after the condensation experiments was calculated to compare the condensation performance of different substrates with patterned wettability. The pattern P2 showed the lowest and the pattern P4 showed the highest condensation removal rate in both environmental conditions, as shown in Fig. \ref{fig:mass}a. The formation of larger-sized liquid bulges observed on both sides of the pattern P2 substrate caused the delay in the condensate transport to the wicking reservoir. However, the wicking occurs when the liquid interface of the liquid bulge crosses the side edge of the substrate. Pattern P4 with hybrid SHL tracks has shown a better condensation removal rate compared to pattern P3 with wedge-shaped SHL tracks.  

The effective condensation heat transfer coefficient (CHTC) of the substrate is defined as \cite{mahapatra2016key},
\begin{equation}
    h=\frac{\dot{m} h_{fg}}{T_{dew}-T_s}
    \label{eq:h}
\end{equation}
where $\dot{m}$ is the total mass collected by the wicking reservoir, $h_{fg}$ is the latent heat absorbed by the substrate during condensation, $T_{dew}$ is the dew point temperature, and $T_s$ is the substrate temperature. The accurate collection of condensate without any mass losses is important for predicting the CHTC accurately. As the mass collection system is kept in a humid environment, evaporation of the collected condensate occurs and can lead to the underprediction of the CHTC. In the present study, condensate collection is achieved with a wicking reservoir, as shown in Fig. \ref{fig:w_res}. Here, the wicking reservoir is covered with a plastic sheet on all the faces except the interior faces parallel to the substrate edges, through which the condensate is absorbed. Hence, evaporation occurs only through the faces that are parallel to the substrate edges. The temperature of the humid air near the interior faces of the wicking reservoir is much lower than the $T_{env}$ due to sensible cooling. Hence, the evaporation loss during condensate collection is negligible and the corresponding contribution to the heat transfer coefficient is not accounted in Eq. \ref{eq:h}.

The formation of a greater number of liquid bulges reduces the CHTC on pattern P2 substrate by $\sim$15\% and $\sim$10\% compared to pattern P1 at the experimental conditions A and B, respectively. The flooding of condensate was not observed in patterns P3 and P4. The passive transport of condensate through SHL wedge tracks on the patterns P2 and P4 avoided the flooding of condensate on the substrate. Hence, the condensation performance on substrates having patterns P3 and P4 was found to be better as compared to the substrates with patterns P1 and P2. In experimental condition A, the CHTC of pattern P3 and P4 substrates was enhanced by $\sim$136\% and $\sim$178\% compared to pattern P1. For the experimental condition B, the enhancement was $\sim$107\% and $\sim$138\%, respectively, for the substrates with pattern P3 and P4. Interestingly, the CHTC of pattern P4 was enhanced by $\sim$18\% and $\sim$15\% compared to pattern P3 at environmental conditions A and B, respectively. The main reasons for this enhancement were the increase in the number of SHL tracks and the decrease in the area occupied by the single SHL track on pattern P4 compared to pattern P3.

\subsection{Design limitations and scalability of the condenser platform}
This study explored the feasibility of a plate-type condenser that can operate continuously in a humid air environment in microgravity. The current work attempted to address challenges related to the removal of condensate from the condenser substrate in a microgravity condition. The condensate is transported from the interior area of the horizontally oriented substrate to the side edges by passive transport in the wedge shaped pattern, and its removal is aided by continuous wicking from the side edges by a wicking reservoir. However, the proposed condenser platform may fail once the wicking reservoir is saturated with the condensate collected from the substrate, which could further result in surface flooding. When the condenser is run continuously for a longer duration or under high condensation flux conditions, the wicking reservoir can saturate with the condensate liquid. Hence, the surrounded wicking reservoir can only act as an intermediate reservoir, and condensate must be continuously extracted from the wicking reservoir by an external system during long-term operation. This can be accomplished by making micrometer-sized holes in the wicking reservoir, with one end closed and the other end attached to a micrometer-sized capillary tube. When the wicking reservoir is filled with the condensate absorbed from the substrate, the holes in the reservoir began to fill. Thereafter, the condensate will be carried by capillary action through the connected capillary tube. This capillary tube may continuously suck the collected condensate from the wicking reservoir.  Furthermore, the condensate from this tube can be pumped using either a suction pump \cite{hasan2006conceptual} or a capillary ejection mechanism \cite{wollman2013new,mehrabian2014auto} in microgravity to a storage tank. 
\begin{figure}
    \centering
    \includegraphics[width=0.8\textwidth]{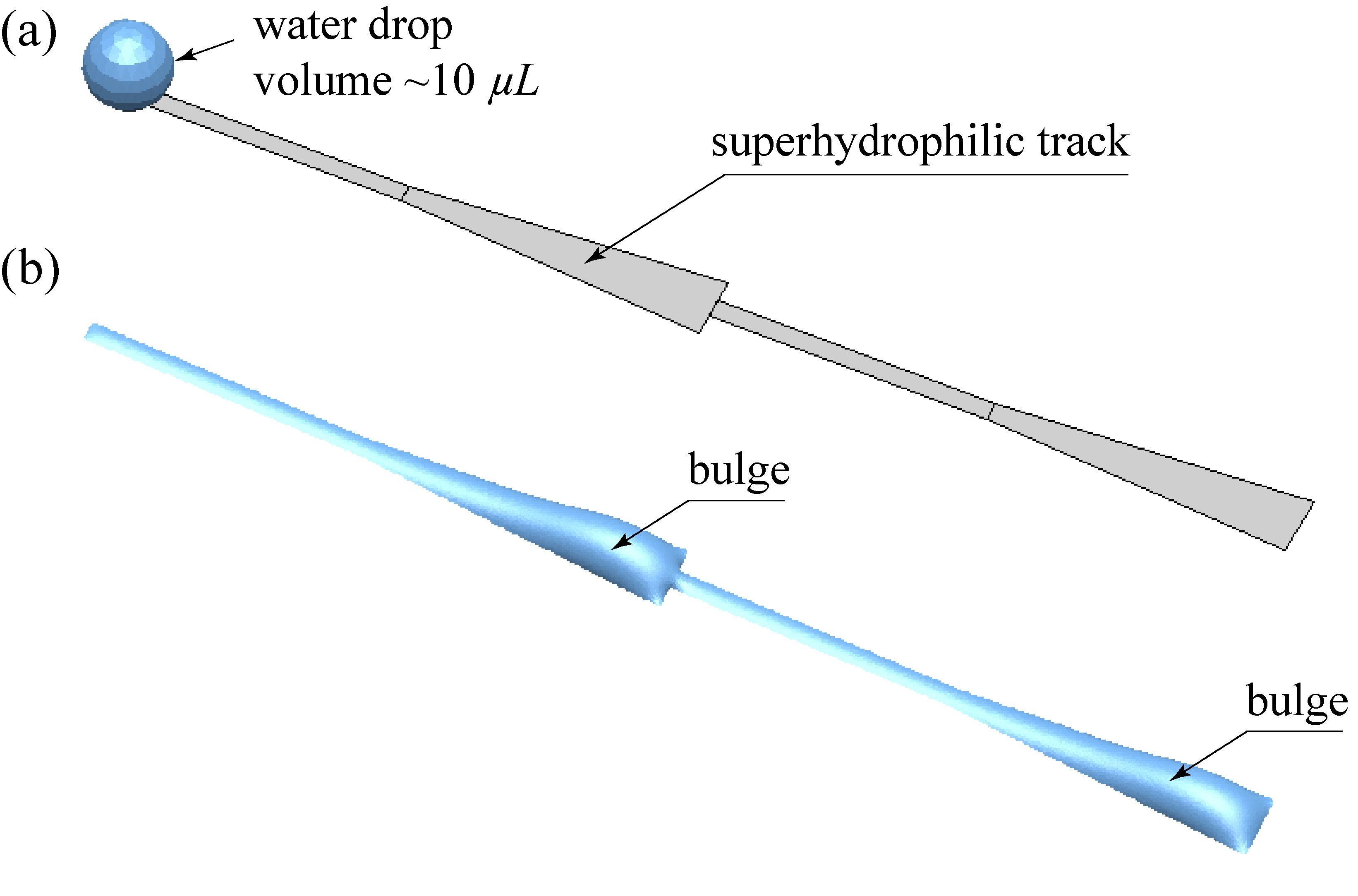}
    \caption{Steady-state simulation of the passive droplet transport in the two superhydrophilic tracks of pattern P4 that have been sandwiched in series. (a) Initial, and (b) final state from the simulation. The volume of the water drop was 10 $\mu L$.}
    \label{fig:longTrack}
\end{figure}

The proposed condenser platform had an active condenser area of 40 $\times$ 40 $mm^2$. During the scaling of the condenser area, the length of the superhydrophilic tracks should also be increased. However, the passive transport of the condensate may fail when the length of the wedge is higher, which can cause surface flooding. Therefore, for a larger substrate, the proposed patterned design must be modified to avoid the flooding issue. In such cases, multiple superhydrophilic tracks in the pattern P4 can be sandwiched in series to increase the overall length of the superhydrophilic track, as shown in Fig. \ref{fig:longTrack}a. The steady-state simulation elucidated that the passive transport of the droplet is possible even with two superhydrophilic tracks that are connected in series (see Fig. \ref{fig:longTrack}b). The formation of a liquid bulge was observed in the wider end of both wedges. The size of the liquid bulges will increase when the droplets from the superhydrophobic region are transported to the superhydrophilic track by drop-film coalescence. Further, the wicking at the side edges will occur when the interface of the liquid bulge close to the side edges exceeds the gap between the substrate and wicking reservoir.

A sufficient wettability contrast between the hydrophilic and hydrophobic regions is a primary requirement for the efficient transport of condensate through the superhydrophilic wedge track and the transport of drop from the hydrophobic region to the hydrophilic track \cite{ghosh2014wettability, yan2022microscale}. The absorption of volatile organic compounds (VOCs) present in the atmosphere might raise the contact angle of the superhydrophilic region in ambient conditions \cite{ma2020recent}. However, such possibilities are rare on the space station because atmospheric air conditions are artificially created in the space station using purified gases carried from Earth. Despite repeated condensation cycles, the superhydrophobic characteristic of the fabricated self assembled mono-layer coating did not change, indicating that the superhydrophobic coating was chemically stable.

In practical applications, the proposed wicking reservoir made from layers of filter papers is inadequate and can be interchanged with metallic wicking materials that offer better durability. Further research is required to develop a working prototype for microgravity conditions that can be operated for long durations or under high heat flux conditions. In such cases, the proposed condenser platform must be integrated with an external system that can continuously transport the collected condensate from the wicking reservoir to a storage tank. The proposed condensate removal mechanism, which utilises patterns of wettability and wicking reservoirs, can also be adopted for high heat flux condensation. This study explored the experiments of water vapour condensation in the presence of air. Therefore, the exhibited technology is applicable to the management of humidity within the space station. After filtration and mineralization, the collected condensate from the platform can also be used as drinking water for astronauts. 

\section{Conclusion}
In this study, we proposed a proof-of-concept for a novel plate-type condenser platform that facilitates the continuous removal of condensate from a horizontal wettability-patterned surface via a wicking reservoir surrounding the condensing area. This study examined the condensation dynamics and heat transfer performance of four distinct patterns of wettability comprised of laser-engraved superhydrophilic tracks of varying shapes on a superhydrophobic surface. Using coalescence-driven capillary pumping, the condensate from the superhydrophobic region migrated to the superhydrophilic tracks. The physics of the passive transport of the condensate liquid on various patterns of wettability is investigated using numerical modelling. The Laplace pressure difference of the wedge-shaped superhydrophilic track directs the condensate to the substrate's side edges. Further, the accumulated condensate along the side edges is absorbed by the wicking reservoir. Flooding of the condensate is observed on the engineered wettability designs of P1 and P2. On the engineered wettability designs P3 and P4, we observed the continuous removal of condensate without flooding. Pattern P4 has demonstrated better condensation heat transfer performance than pattern P3 by more than 15\%. Thus, the optimal surface with a patterned wettability can enhance the effective condensation performance. The current research could pave the way for developing efficient condenser surfaces for space applications.      

\section*{Acknowledgement}
The authors thank Science and Engineering Research Board (SERB) of Government of India [Project number: ECR/2018/001806] for funding the work. This research is partly supported by Indian Institute of Technology Madras to the Micro Nano Bio Fluidics Group under the funding for Institutions of Eminence scheme of Ministry of Education, Government of India [Sanction. No: $11/9/2019-U.3(A)$].

\bibliographystyle{unsrt} 
\bibliography{ass-refs}

\begin{thebibliography}{10}

\bibitem{li2019organic}
Xiaoya Li, Jian Song, Guopeng Yu, Youcai Liang, Hua Tian, Gequn Shu, and
  Christos~N Markides.
\newblock Organic rankine cycle systems for engine waste-heat recovery: Heat
  exchanger design in space-constrained applications.
\newblock {\em Energy Conversion and Management}, 199:111968, 2019.

\bibitem{perez2011review}
Luis P{\'e}rez-Lombard, Jos{\'e} Ortiz, Juan~F Coronel, and Ismael~R Maestre.
\newblock A review of hvac systems requirements in building energy regulations.
\newblock {\em Energy and Buildings}, 43(2-3):255--268, 2011.

\bibitem{nioras2021different}
Dimitrios Nioras, Kosmas Ellinas, Vassilios Constantoudis, and Evangelos
  Gogolides.
\newblock How different are fog collection and dew water harvesting on surfaces
  with different wetting behaviors?
\newblock {\em ACS Applied Materials \& Interfaces}, 13(40):48322--48332, 2021.

\bibitem{nassrullah2020energy}
Haya Nassrullah, Shaheen~Fatima Anis, Raed Hashaikeh, and Nidal Hilal.
\newblock Energy for desalination: A state-of-the-art review.
\newblock {\em Desalination}, 491:114569, 2020.

\bibitem{song2021multifunctional}
Ying-Nan Song, Mao-Qin Lei, Dong-Lin Han, Yu-Chuan Huang, Shuai-Peng Wang,
  Jian-Yang Shi, Yue Li, Ling Xu, Jun Lei, and Zhong-Ming Li.
\newblock Multifunctional membrane for thermal management applications.
\newblock {\em ACS Applied Materials \& Interfaces}, 13(16):19301--19311, 2021.

\bibitem{schmidt1930condensation}
E~Schmidt, W~Schurig, and W~Sellschopp.
\newblock Condensation of water vapour in film-and drop form.
\newblock {\em Technische Mechanik und Thermodynamik}, 1:53--63, 1930.

\bibitem{tanaka1979further}
H~Tanaka.
\newblock Further developments of dropwise condensation theory.
\newblock {\em Journal of Heat Transfer}, 101:603--611, 1979.

\bibitem{attinger2014surface}
Daniel Attinger, Christophe Frankiewicz, Amy~R Betz, Thomas~M Schutzius, Ranjan
  Ganguly, Arindam Das, Chang-Jin Kim, and Constantine~M Megaridis.
\newblock Surface engineering for phase change heat transfer: A review.
\newblock {\em MRS Energy \& Sustainability}, 1, 2014.

\bibitem{cho2016nanoengineered}
H~Jeremy Cho, Daniel~J Preston, Yangying Zhu, and Evelyn~N Wang.
\newblock Nanoengineered materials for liquid--vapour phase-change heat
  transfer.
\newblock {\em Nature Reviews Materials}, 2(2):1--17, 2016.

\bibitem{tran2022ultrafast}
Ngoc~Giang Tran and Doo-Man Chun.
\newblock Ultrafast and eco-friendly fabrication process for robust, repairable
  superhydrophobic metallic surfaces with tunable water adhesion.
\newblock {\em ACS Applied Materials \& Interfaces}, 2022.

\bibitem{wilke2020polymer}
Kyle~L Wilke, Dion~S Antao, Samuel Cruz, Ryuichi Iwata, Yajing Zhao, Arny
  Leroy, Daniel~J Preston, and Evelyn~N Wang.
\newblock Polymer infused porous surfaces for robust, thermally conductive,
  self-healing coatings for dropwise condensation.
\newblock {\em ACS Nano}, 14(11):14878--14886, 2020.

\bibitem{mangini2017hybrid}
Daniele Mangini, Mauro Mameli, Davide Fioriti, Sauro Filippeschi, L~Araneo, and
  Marco Marengo.
\newblock Hybrid pulsating heat pipe for space applications with non-uniform
  heating patterns: ground and microgravity experiments.
\newblock {\em Applied Thermal Engineering}, 126:1029--1043, 2017.

\bibitem{chatterjee2011constrained}
Arya Chatterjee, Peter~C Wayner~Jr, Joel~L Plawsky, David~F Chao, Ronald~J
  Sicker, Tibor Lorik, Louis Chestney, John Eustace, Raymond Margie, and John
  Zoldak.
\newblock The constrained vapor bubble fin heat pipe in microgravity.
\newblock {\em Industrial \& Engineering Chemistry Research},
  50(15):8917--8926, 2011.

\bibitem{hasan2006conceptual}
Mohammad~M Hasan, Lutful~I Khan, Vedha Nayagam, and Ramaswamy Balasubramaniam.
\newblock Conceptual design of a condensing heat exchanger for space systems
  using porous media.
\newblock In {\em 35th International Conference on Environmental Systems
  (ICES)}, pages E--15464, 2006.

\bibitem{hauser2002condensing}
Gerhard Hauser, Ludwig Eicher, Joachim Lucas, Johannes Witt, and Tanguy Morel.
\newblock Condensing heat exchanger, July~16 2002.
\newblock \uppercase{US} Patent 6,418,743.

\bibitem{thomas2014condensing}
Christopher~M Thomas and Yonghui Ma.
\newblock Condensing heat exchanger with hydrophilic antimicrobial coating,
  July~1 2014.
\newblock \uppercase{US} Patent 8,763,682.

\bibitem{lowrey2021survey}
Sam Lowrey, Kirill Misiiuk, Richard Blaikie, and Andrew Sommers.
\newblock Survey of micro/nanofabricated chemical, topographical, and compound
  passive wetting gradient surfaces.
\newblock {\em Langmuir}, 38(2):605--619, 2021.

\bibitem{dai2020directional}
Haoyu Dai, Zhichao Dong, and Lei Jiang.
\newblock Directional liquid dynamics of interfaces with superwettability.
\newblock {\em Science Advances}, 6(37):eabb5528, 2020.

\bibitem{thomas2021droplet}
Tibin~M Thomas, Imdad~Uddin Chowdhury, K~Dhivyaraja, Pallab~Sinha Mahapatra,
  Arvind Pattamatta, and Manish~K Tiwari.
\newblock Droplet dynamics on a wettability patterned surface during spray
  impact.
\newblock {\em Processes}, 9(3):555, 2021.

\bibitem{sinha2022patterning}
Pallab Sinha~Mahapatra, Ranjan Ganguly, Aritra Ghosh, Souvick Chatterjee, Sam
  Lowrey, Andrew~D Sommers, and Constantine~M Megaridis.
\newblock Patterning wettability for open-surface fluidic manipulation:
  Fundamentals and applications.
\newblock {\em Chemical Reviews}, 122(22):16752--16801, 2022.

\bibitem{chowdhury2019self}
Imdad~Uddin Chowdhury, Pallab Sinha~Mahapatra, and Ashis~Kumar Sen.
\newblock Self-driven droplet transport: Effect of wettability gradient and
  confinement.
\newblock {\em Physics of Fluids}, 31(4):042111, 2019.

\bibitem{sen2018scaling}
Uddalok Sen, Souvick Chatterjee, Ranjan Ganguly, Richard Dodge, Lisha Yu, and
  Constantine~M Megaridis.
\newblock Scaling laws in directional spreading of droplets on
  wettability-confined diverging tracks.
\newblock {\em Langmuir}, 34(5):1899--1907, 2018.

\bibitem{li2017topological}
Jiaqian Li, Xiaofeng Zhou, Jing Li, Lufeng Che, Jun Yao, Glen McHale, Manoj~K
  Chaudhury, and Zuankai Wang.
\newblock Topological liquid diode.
\newblock {\em Science Advances}, 3(10):eaao3530, 2017.

\bibitem{sun2019surface}
Qiangqiang Sun, Dehui Wang, Yanan Li, Jiahui Zhang, Shuji Ye, Jiaxi Cui,
  Longquan Chen, Zuankai Wang, Hans-J{\"u}rgen Butt, Doris Vollmer, and
  Xu~Deng.
\newblock Surface charge printing for programmed droplet transport.
\newblock {\em Nature Materials}, 18(9):936--941, 2019.

\bibitem{stamatopoulos2020droplet}
Christos Stamatopoulos, Athanasios Milionis, Norbert Ackerl, Matteo Donati,
  Paul Leudet de~la Vallee, Philipp Rudolf~von Rohr, and Dimos Poulikakos.
\newblock Droplet self-propulsion on superhydrophobic microtracks.
\newblock {\em ACS Nano}, 14(10):12895--12904, 2020.

\bibitem{zhang2021charge}
Chenglin Zhang, Dehui Wang, Jinlong Yang, Wenluan Zhang, Qiangqiang Sun, Fanfei
  Yu, Yue Fan, Yong Li, Longquan Chen, and Xu~Deng.
\newblock Charge density gradient propelled ultrafast sweeping removal of
  dropwise condensates.
\newblock {\em The Journal of Physical Chemistry B}, 125(7):1936--1943, 2021.

\bibitem{boreyko2009self}
Jonathan~B Boreyko and Chuan-Hua Chen.
\newblock Self-propelled dropwise condensate on superhydrophobic surfaces.
\newblock {\em Physical Review Letters}, 103(18):184501, 2009.

\bibitem{aili2016unidirectional}
Abulimiti Aili, Hongxia Li, Mohamed~H Alhosani, and TieJun Zhang.
\newblock Unidirectional fast growth and forced jumping of stretched droplets
  on nanostructured microporous surfaces.
\newblock {\em ACS Applied Materials \& Interfaces}, 8(33):21776--21786, 2016.

\bibitem{preston2018gravitationally}
Daniel~J Preston, Kyle~L Wilke, Zhengmao Lu, Samuel~S Cruz, Yajing Zhao,
  Laura~L Becerra, and Evelyn~N Wang.
\newblock Gravitationally driven wicking for enhanced condensation heat
  transfer.
\newblock {\em Langmuir}, 34(15):4658--4664, 2018.

\bibitem{cheng2021rapid}
Yaqi Cheng, Mingmei Wang, Jing Sun, Minjie Liu, Bingang Du, Yuanbo Liu, Yuankai
  Jin, Rongfu Wen, Zhong Lan, Xiaofeng Zhou, et~al.
\newblock Rapid and persistent suction condensation on hydrophilic surfaces for
  high-efficiency water collection.
\newblock {\em Nano Letters}, 21(17):7411--7418, 2021.

\bibitem{bahadur2007electrowetting}
Vaibhav Bahadur and Suresh~V Garimella.
\newblock Electrowetting-based control of static droplet states on rough
  surfaces.
\newblock {\em Langmuir}, 23(9):4918--4924, 2007.

\bibitem{daniel2001fast}
Susan Daniel, Manoj~K Chaudhury, and John~C Chen.
\newblock Fast drop movements resulting from the phase change on a gradient
  surface.
\newblock {\em Science}, 291(5504):633--636, 2001.

\bibitem{macner2014condensation}
Ashley~M Macner, Susan Daniel, and Paul~H Steen.
\newblock Condensation on surface energy gradient shifts drop size distribution
  toward small drops.
\newblock {\em Langmuir}, 30(7):1788--1798, 2014.

\bibitem{sharma2018gladiolus}
Vipul Sharma, Daniel Orejon, Yasuyuki Takata, Venkata Krishnan, and
  Sivasankaran Harish.
\newblock Gladiolus dalenii based bioinspired structured surface via soft
  lithography and its application in water vapor condensation and fog
  harvesting.
\newblock {\em ACS Sustainable Chemistry \& Engineering}, 6(5):6981--6993,
  2018.

\bibitem{zhang2017bioinspired}
Songnan Zhang, Jianying Huang, Zhong Chen, and Yuekun Lai.
\newblock Bioinspired special wettability surfaces: from fundamental research
  to water harvesting applications.
\newblock {\em Small}, 13(3):1602992, 2017.

\bibitem{park2016condensation}
Kyoo-Chul Park, Philseok Kim, Alison Grinthal, Neil He, David Fox, James~C
  Weaver, and Joanna Aizenberg.
\newblock Condensation on slippery asymmetric bumps.
\newblock {\em Nature}, 531(7592):78--82, 2016.

\bibitem{hou2015recurrent}
Youmin Hou, Miao Yu, Xuemei Chen, Zuankai Wang, and Shuhuai Yao.
\newblock Recurrent filmwise and dropwise condensation on a beetle mimetic
  surface.
\newblock {\em ACS Nano}, 9(1):71--81, 2015.

\bibitem{sun2019microdroplet}
Jianxing Sun and Patricia~B Weisensee.
\newblock Microdroplet self-propulsion during dropwise condensation on
  lubricant-infused surfaces.
\newblock {\em Soft Matter}, 15(24):4808--4817, 2019.

\bibitem{ghosh2014enhancing}
Aritra Ghosh, Sara Beaini, Bong~June Zhang, Ranjan Ganguly, and Constantine~M
  Megaridis.
\newblock Enhancing dropwise condensation through bioinspired wettability
  patterning.
\newblock {\em Langmuir}, 30(43):13103--13115, 2014.

\bibitem{mahapatra2016key}
Pallab~Sinha Mahapatra, Aritra Ghosh, Ranjan Ganguly, and Constantine~M
  Megaridis.
\newblock Key design and operating parameters for enhancing dropwise
  condensation through wettability patterning.
\newblock {\em International Journal of Heat and Mass Transfer}, 92:877--883,
  2016.

\bibitem{nioras2022atmospheric}
Dimitrios Nioras, Kosmas Ellinas, and Evangelos Gogolides.
\newblock Atmospheric water harvesting on micro-nanotextured biphilic surfaces.
\newblock {\em ACS Applied Nano Materials}, 2022.

\bibitem{tang2021design}
Yu~Tang, Xiaolong Yang, Yimin Li, and Di~Zhu.
\newblock Design of hybrid superwetting surfaces with self-driven droplet
  transport feature for enhanced condensation.
\newblock {\em Advanced Materials Interfaces}, 8(13):2100284, 2021.

\bibitem{koukoravas2020experimental}
Theodore~P Koukoravas, George Damoulakis, and Constantine~M Megaridis.
\newblock Experimental investigation of a vapor chamber featuring
  wettability-patterned surfaces.
\newblock {\em Applied Thermal Engineering}, 178:115522, 2020.

\bibitem{derby2014flow}
Melanie~M Derby, Abhra Chatterjee, Yoav Peles, and Michael~K Jensen.
\newblock Flow condensation heat transfer enhancement in a mini-channel with
  hydrophobic and hydrophilic patterns.
\newblock {\em International Journal of Heat and Mass Transfer}, 68:151--160,
  2014.

\bibitem{vedder1969aluminum}
W~Vedder and DA~Vermilyea.
\newblock Aluminum+water reaction.
\newblock {\em Transactions of the Faraday Society}, 65:561--584, 1969.

\bibitem{thomas2021condensation}
Tibin~M Thomas and Pallab Sinha~Mahapatra.
\newblock Condensation of humid air on superhydrophobic surfaces: Effect of
  nanocoatings on a hierarchical interface.
\newblock {\em Langmuir}, 37(44):12767--12780, 2021.

\bibitem{gao2009wetting}
Lichao Gao and Thomas~J McCarthy.
\newblock Wetting 101°.
\newblock {\em Langmuir}, 25(24):14105--14115, 2009.

\bibitem{yang2006dropwise}
Qingfeng Yang and Anzhong Gu.
\newblock Dropwise condensation on sam and electroless composite coating
  surfaces.
\newblock {\em Journal of Chemical Engineering of Japan}, 39(8):826--830, 2006.

\bibitem{wen2017hierarchical}
Rongfu Wen, Shanshan Xu, Dongliang Zhao, Yung-Cheng Lee, Xuehu Ma, and Ronggui
  Yang.
\newblock Hierarchical superhydrophobic surfaces with micropatterned nanowire
  arrays for high-efficiency jumping droplet condensation.
\newblock {\em ACS Applied Materials \& Interfaces}, 9(51):44911--44921, 2017.

\bibitem{matsui2012hybrid}
Hiroyuki Matsui, Yuki Noda, and Tatsuo Hasegawa.
\newblock Hybrid energy-minimization simulation of equilibrium droplet shapes
  on hydrophilic/hydrophobic patterned surfaces.
\newblock {\em Langmuir}, 28(44):15450--15453, 2012.

\bibitem{ghosh2014wettability}
Aritra Ghosh, Ranjan Ganguly, Thomas~M Schutzius, and Constantine~M Megaridis.
\newblock Wettability patterning for high-rate, pumpless fluid transport on
  open, non-planar microfluidic platforms.
\newblock {\em Lab on a Chip}, 14(9):1538--1550, 2014.

\bibitem{wollman2013new}
Andrew Wollman and Mark Weislogel.
\newblock New investigations in capillary fluidics using a drop tower.
\newblock {\em Experiments in Fluids}, 54(4):1--13, 2013.

\bibitem{mehrabian2014auto}
Hadi Mehrabian and James~J Feng.
\newblock Auto-ejection of liquid drops from capillary tubes.
\newblock {\em Journal of Fluid Mechanics}, 752:670--692, 2014.

\bibitem{yan2022microscale}
Xiao Yan, Feipeng Chen, Chongyan Zhao, Xiong Wang, Longnan Li, Siavash
  Khodakarami, Kazi Fazle~Rabbi, Jiaqi Li, Muhammad~Jahidul Hoque, Feng Chen,
  et~al.
\newblock Microscale confinement and wetting contrast enable enhanced and
  tunable condensation.
\newblock {\em ACS Nano}, 16(6):9510--9522, 2022.

\bibitem{ma2020recent}
Jingcheng Ma, Soumyadip Sett, Hyeongyun Cha, Xiao Yan, and Nenad Miljkovic.
\newblock Recent developments, challenges, and pathways to stable dropwise
  condensation: a perspective.
\newblock {\em Applied Physics Letters}, 116(26):260501, 2020.

\end{thebibliography}

\end{document}